# Sitnikov in Westeros: How Celestial Mechanics finally explains why winter is coming in Game of Thrones

**Florian Freistetter**[1]
*die siebente fakultät: Zentrum für Gesellschaft, Wissen und Kommunikation*
*Karl-Franzens-Universität Graz*
*Schubertstrasse 1/EG*
*8010 Graz*
*Austria*

**Ruth Grützbauch**[2]
*Public Space Pop-up Planetarium*
*Friedrich-Kaiser-Gasse 37/6*
*1160 Wien*
*Austria*

**Abstract**

*"Winter is coming". As far as meteorological predictions go, the words of House Stark are both trivial and not very helpful for a scientific analysis to explain the chaotic sequence of the seasons in the world of "Game of Thrones". The natives of Westeros have failed to develop a feasible model to understand and predict the coming and duration of their winters. And although the scientists of Earth have brought forth many different mechanisms to explain the seasons, all of them are found wanting (at least by us). Nobody seems to have discovered the one and only true and working mechanism to explain the coming and going of winters on Westeros and thus it is up to us to introduce the world to the might of the often ignored Sitnikov problem. That very special configuration of two stars and one planet is ideally suited to explain everything that needs to be explained and uncover the real reason for the coming of winter.*
[submitted on April 1, 2018]

## 1. Introduction

"Winter is coming", as the words of House Stark have been proclaiming for centuries. But nobody seems to know when the cold season will make its appearance and how long it will last, when it finally comes. The meteorological knowledge of the people of Westeros seems to be unreasonably non-existent, considering the importance of the seasons. After all, winter in Westeros brings not only snow, black ice and the common cold, but holds the potential for zombie invasions, ice dragons and the end of the world as we know it. One should think that the learned maesters of the citadel had developed some knowledge in the fields of climatology and celestial mechanics to identify the source of their home world's erratic seasons and find ways to predict the coming of winter. However, as far as we know, no such sciences do exist in Westeros and it therefore falls to the scientists of Earth to explain what is going on with the seasons in the world of "Game of Thrones".

---
[1] florian@extrasolar-planets.de - http://scienceblogs.de/astrodicticum-simplex
[2] hello@publicspace.at - http://publicspace.at

Seasons as such are not difficult to explain. On Earth they are the result of the Earth's axial tilt relative to the plane in which our planet orbits the sun. During the course of one year, different hemispheres thus receive sunlight with different intensities and for different amounts of time. On Westeros, the whole situation is much more complicated: the seasonal patterns do not seem to coincide with the yearly orbit of the planet around the star. Winter can come anytime and it can stay for any length of time. This requires a much more sophisticated explanation than just an axial tilt[3].

Luckily, astronomers have known about just such an explanation for more than half a century (at least those astronomers, who do not waste their time looking up to the sky but follow the noble path of pure theory and mathematical analysis of the universe[4]). It is called the "Sitnikov problem" and is able to explain everything that needs to be explained concerning the seasons of Westeros.

**2. Proposals that almost seem to work but in fact don't:**

Explaining the seasons of the world of "Game of Thrones" is a longstanding problem in the field of planetary science of fantasy planets (admittedly a very small and not yet fully embraced field of research) and many people have tried to solve it. Tried, and failed, as the black brothers of the Night's Watch have tried and failed to protect Westeros from the onslaught of White Walkers and the army of the dead. We list the failed attempts in the following sections, but since they are all insufficient in explaining the seasons of Westeros, the reader is encouraged to skip the whole part and go straight to our cool and new (and working!) explanation presented in section 3.

**2.1. Binary Stars**

Kostov et al (2013) proposed a binary star system, in which a planet orbits two solar like stars. Such circumbinary planets (see e.g. Welsh et al. 2012) do exist in reality, but they are not able to explain all features of the erratic seasons that happen on Westeros. First, there is the problem of dynamical stability. There are many regions of stable planetary motion in double star systems (e.g. Pilat-Lohinger et al. 2002, Dvorak 1984), but a planet must either orbit one of the two stars in a very close orbit or far outside around both of the stars. In the first case, the second star in general has to be so far away that no significant influence on the climate of the planet is expected. In the second case, the planet is far enough from both stars, so that it receives their combined energy as if coming from a single star. To get seasonal changes like on Westeros, a planet has to orbit in the vicinity of both stars, which is a dynamically unstable situation. Such a planet follows a stable motion for a short time, but will ultimately collide with one of the stars or be ejected of the system. Especially, there will be not enough time for life to evolve on such a planet, and as well as for all the background story of "Game of Thrones" to take place. Although Kostov et al. (2013) showed in their work that one can get a chaotic sequence of seasons on a circumbinary planet, they investigated the motion of the planet only for an integration time of one million days which is of course way to short for any meaningful analysis of planetary stability.

---

[3] Either that, or it's just magic. But although magic is quite fun to read about in a fantasy book, it is much less fun for a scientist, since it can explain everything and thus puts us all out of a job.
[4] The second author explicitly disagrees with this statement.

An additional problem exists with this kind of explanation. The astronomical knowledge of the people of Westeros may leave a lot to be desired. But one would expect them to notice the difference between one and two suns in their sky. Since the model of Kostov et al. (2013) needs two solar like stars to produce the chaotic sequence of seasons, people should have noticed the two suns over their heads. Since we never hear about a second sun[5] on Westeros, we can discard this hypothesis and similar proposals (Merrifield 2014).

## 2.2. Wobbly axial tilt

Since the seasons on Earth are caused by the planet's axial tilt, the simplest method to explain the seasons on Westeros seems to be to tinker with its axis too (which is why this mechanism has been proposed by many: Griffith & Douglas 2017, Selcke 2015, Merrifield 2014, Dvorsky 2012). Our seasons follow a stable pattern because Earth's axial tilt is stable and does not deviate much from its value of 23.5 degrees. If the variations would be bigger, the seasonal cycle would no longer be constant. We know that the tilt of Earth's axis is stabilised by the gravity of our Moon (Laskar 1993). However, without its presence, there probably would be no stable long term climate on Earth. But we know that there is a moon in the sky of Westeros and although we do not know how big it is (again that deplorable lack of astronomical knowledge!), it is at least big enough to be seen with the naked eye and to give a noticeable amount of light during the night. It thus should be able to exercise a stabilising influence like our Moon. Even if there is no moon over Westeros to stabilise its axis, we know from numerical models that changes in the axial tilt are very slow and we would not expect any short time effects on the planet's seasons. To get the short time variation of the axial tilt needed to explain the seasons on Westeros, huge amounts of energy would have to be put into the system. For instance, hitting the planet with another planet would deliver the required energy, but would probably also destroy Westeros. And although a global mass extinction event would fit the general theme of George R.R. Martin killing everybody, it is neither of interesting scientific value, nor is it much fun.

## 2.3. Milankovic Cycles

Proposing Milankovic Cycles as the cause for the seasons of Westeros (Griffith & Douglas 2017, Dvorsky 2012) has similar problems of time scales as the modification of the planet's axis described above. Dynamical simulations of the motion of planets in the Solar system show that Earth's orbit is subject to many perturbations by the gravitational pull of the other planets. Over very long time scales these perturbations can cause warm ages and ice ages. To produce not ages but seasons, the perturbations of the planets have to act on much, much shorter time scales which in turn requires very high energies and such planetary systems would probably no longer be stable.

---

[5] Not to be confused with the sellsword company of the "Second Sons", who have nothing to do with astronomy.

## 2.4. Eccentric Orbit

Some authors (Dvorsky 2012, Merrifield 2014) propose, that the planet of Westeros follows a very eccentric orbit around its star. In that case, one would get short summers if the planet is close to the star and long winters if it approaches the apastron of the orbit. Planets on such eccentric orbits can exist, but although the seasons in that case can be very different from the seasons on Earth, they would still follow a regular pattern. And even though the maesters of Westeros do not know much about astronomy, they should have been able to spot such a pattern and develop reasonable predictions concerning the coming of Winter.

## 2.5. Volcanic Activity & Oceanic Currents

Some authors have refused astronomy and proposed a planetary origin of the chaotic seasons on Westeros. They claim, that volcanic activity (Griffith & Douglas 2017, Dvorsky 2012) or changing currents in the ocean (Dvorsky 2012, Merrifield 2014) can cause changes in the planet's climate. We know from Earth that volcanic eruptions can trigger large scale temperature changes (like the "year without summer" that followed the eruption of the Tambora in 1815) and that currents such as the gulf stream have a big influence on the distribution of heat around the planet. But if Westeros is plagued for millennia by the eruptions of giant volcanoes, one would think that somebody would have noticed. The same holds true for the ocean currents, that surly are know to seafaring people like the ironborn of House Greyjoy.

## 2.6. Variable Stars

Merrifield (2014) proposed a very simple solution for the problem of the chaotic seasons. The planet of Westeros could orbit a variable star, that delivers varying amounts of energy to its planet. This would work, but planets around such stars are known to be bad places for the development of life. They often produce bursts of gamma and x-rays, which would either destroy any developing lifeforms or hinder the evolution of life in the first place.

## 2.7. Asteroids

Griffith & Douglas (2017) put the world of Westeros in a planetary system with a very dense asteroid belt and on an eccentric orbit that crosses this belt. This would enhance the number of small bodies that collide with the planet and - as with a volcanic eruption - every asteroid impact big enough could trigger a planet wide drop of temperatures. The crossing of the belt would be periodic, the following impacts however do not have to be periodic themselves, and the consequences of each individual impact would depend on the size of the respective asteroid. Such a configuration would indeed be a very good explanation of the chaotic seasons on Westeros. But as with the volcanos, one would expect the people of Westeros to notice if their homeworld gets smashed by giant celestial bodies over millenia. Since they do not, we have to dispose of this elegant, but still insufficient mechanism.

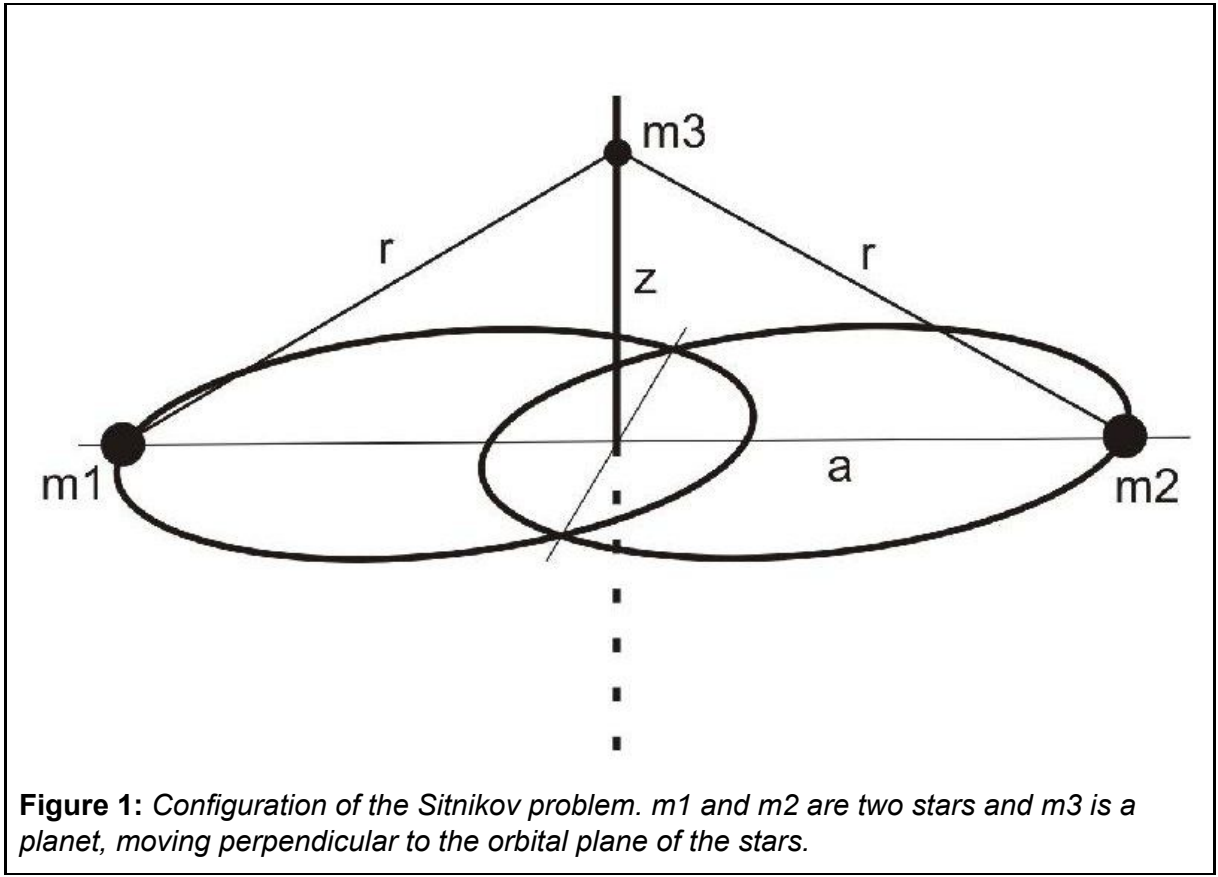

**Figure 1:** *Configuration of the Sitnikov problem. m1 and m2 are two stars and m3 is a planet, moving perpendicular to the orbital plane of the stars.*

### 3. The Sitnikov problem

Now that we know which explanations do not work, we want to introduce the one astronomical configuration which is able to successfully account for the chaotic seasons of Westeros: the "Sitnikov Problem" (Sitnikov 1961). It is a sub-case of the elliptical, restricted three-body problem. Due to the complexity of the possible kinds of motions in this system, it is - despite its simple mathematical formulation - a well and long studied, unique dynamical model in celestial mechanics.

Let there be three masses $m_1$, $m_2$ and $m_3$, where $m_1$ and $m_2$ are called "primaries" (i.e. the stars) and $m_3$ is a smaller (and in the mathematical model massless) body (i.e. a planet). If we let the two stars orbit under their mutual gravitational influence, they will both move around their common barycenter. Now, let us assume a coordinate system that has its origin in that center of mass. If we put the planet exactly in the barycenter, it will be influenced by the gravity of the two stars (but because it is thought as massless, it itself will not disturb the motion of the stars) and follow an orbit along a straight line through the center of mass and perpendicular to the barycenter of the stars (see Figure 1).

## 3.1. Mathematical explanation

If *z* is the planet's distance from the barycenter for a given time *t*, *G* the gravitational constant, *r* the distance of the stars from the center of mass, and $m_1=m_2=m$, then the equations of motion of the planet are given by:

$$\ddot{z} + \frac{2Gmz}{(r^2 + z^2)^{\frac{3}{2}}} = 0$$

There exists an integrable case of the Sitnikov-Problem, if the orbits of the stars are circular. In that case, the problem is called "MacMillan problem" (MacMillan 1911); but the configuration can also be investigated in a much more general case, where the stars have different masses and follow elliptical orbits. In all cases however, the motion of the planet is taking place in the vicinity of this straight line that intersects the center of gravity of the system perpendicular to the primaries' orbital plane. The planet will oscillate above and below the orbital plane of the stars. This motion can be periodic but depending on the total energy of the system, is strongly chaotic in most of the cases.

For a detailed discussion of the dynamics of the Sitnikov problem see Wodnar (1993), Dvorak & Freistetter (2005), Dvorak & Lhotka (2013), Dvorak & Lhotka (2014) or Lhotka (2015). We really urge you to follow that advice, because the problem is a fascinating case of celestial mechanics and should be studied by every astronomer. But concerning the topic of chaotic seasons, we want to point out one very remarkable property of the Sitnikov problem, discovered by, and cited after Moser (1973):

*Consider a solution of z(t) with infinitely many zeroes $t_k$ (k=0, ±1, ±2, …) which are ordered according to size, $t_k < t_{k+1}$, with $z(t_k)=0$. Then we introduce the integers:*

$$s_k = \left[\frac{t_{k+1} - t_k}{2\pi}\right]$$

*which measures the number of complete revolutions of the primaries between two consecutive zeroes of z(t). This way we can associate to every such solution a double infinite sequence of integers. The main result can be expressed as the converse statement:*

*Theorem: Given a sufficiently small eccentricity ε > 0 there exists an integer m=m(ε) such that any sequence $s_k$ with $s_k \geq m$ corresponds to a solution.*

The theorem has also been generalized to non-zero mass of the third body and for any value of eccentricity (Alekseev 1968a, 1968b, 1969).

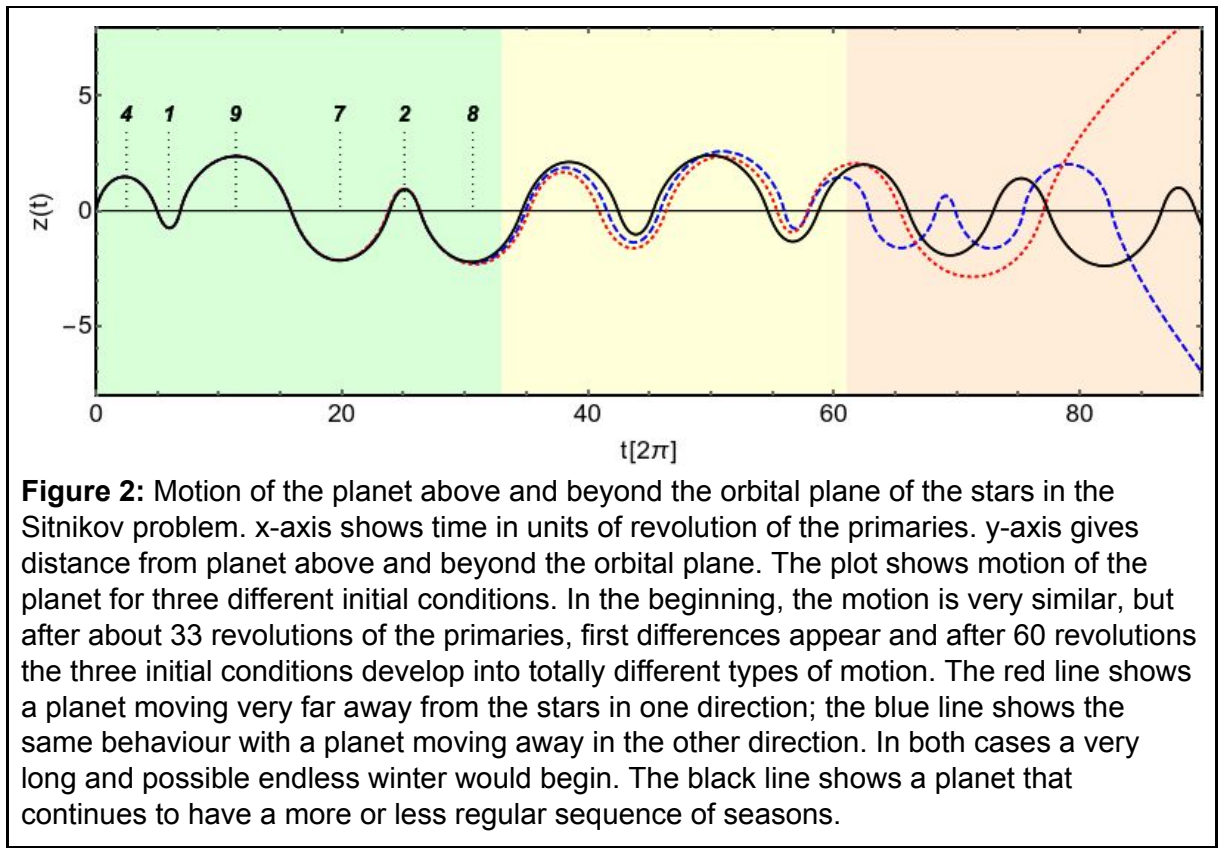

**Figure 2:** Motion of the planet above and beyond the orbital plane of the stars in the Sitnikov problem. x-axis shows time in units of revolution of the primaries. y-axis gives distance from planet above and beyond the orbital plane. The plot shows motion of the planet for three different initial conditions. In the beginning, the motion is very similar, but after about 33 revolutions of the primaries, first differences appear and after 60 revolutions the three initial conditions develop into totally different types of motion. The red line shows a planet moving very far away from the stars in one direction; the blue line shows the same behaviour with a planet moving away in the other direction. In both cases a very long and possible endless winter would begin. The black line shows a planet that continues to have a more or less regular sequence of seasons.

### 3.2. Non-mathematical explanation

For those who - like the maesters of Westeros - refuse to indulge in the wonders of celestial mechanics, let us explain the main point in simpler words. Let's define the time the two stars need for one revolution around the barycenter as one "year". We can then define any infinitely long and arbitrary sequence of whole numbers. The theorem of Mosers stated above ensures that there will be - with absolute certainty - an initial configuration of the two stars and the planet such that the oscillatory motion of the planet will exactly reproduce the chosen sequence of integers. If, for example, we chose the sequence 4,1,9,7,2,8,... then there will be a configuration, where the planet takes four years to move up from the barycenter and down again. It will then spend one year below the orbital plane of the stars before crossing the barycenter again and spending the next nine years above it. Then follows a seven year period below the orbital plane, two years above, eight years below, and so on.

Due to the nonlinear properties of the whole system, very small changes in the initial configuration can produce completely different results (see Figure 2). In fact, this property of chaotic dynamical systems ensures the existence of any other integer sequence in the vicinity of this sequence.

This works for EVERY sequence of integers and that is the key argument to explain the chaotic seasons of Westeros.

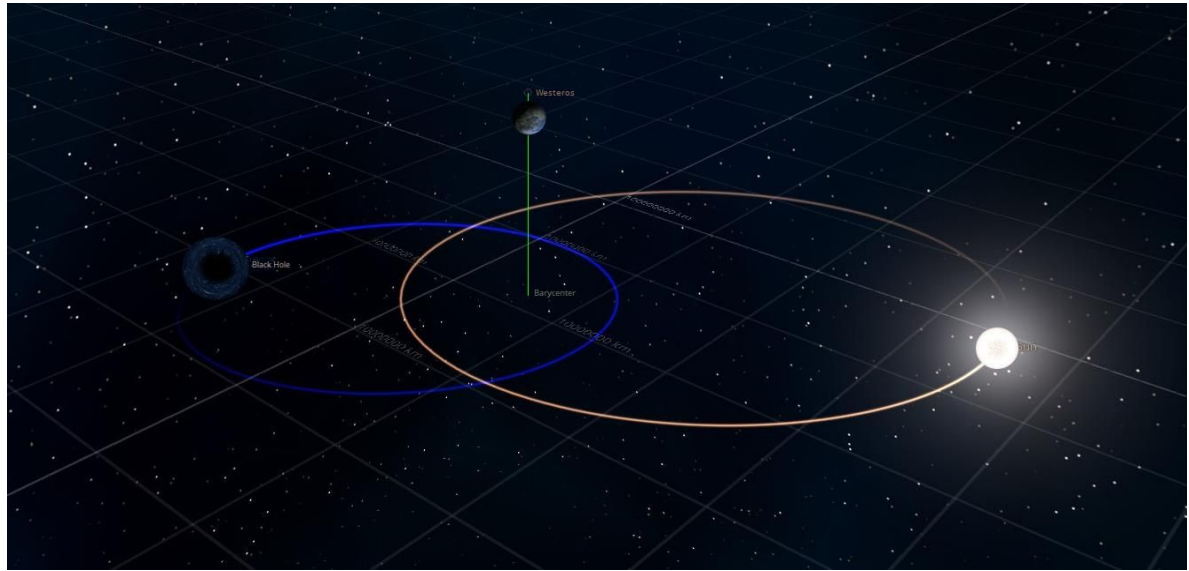

**Figure 3:** Visualization of how a Sitnikov system could look like. The planet oscillates between a star and a black hole.
(Created with Universe Sandbox² http://universesandbox.com/)

### 4. A universal mechanism for every sequence of seasons

Close to the orbital plane of the stars the planet in the Sitnikov configuration will receive a large amount of stellar energy. If the planet is far from the barycenter,then the energy flux will be reduced. In other words: the closer the planet is to the orbital plane of the stars, the warmer it will be. A planet that oscillates regularly above and beyond the plane can have regular seasons like on Earth. But as the theorem of Mosers shows, a Sitnikov-planet can oscillate with any sequence of durations possible. The farther away it is from the plane, the longer the cold season will be. To reproduce a seemingly chaotic sequence of summers and winters of different lengths, one has just to choose an integer sequence accordingly and the theorem ensures, that there will be a configuration of the stars and the planet, that follows exactly that order of seasons.

*However complicated and chaotic the seasons of Westeros are (or will be in the future of "Game of Thrones"): one can find a planet moving in a Sitnikov configuration that has exactly that type of seasons.*

Some people now might want to argue that there are two stars in this configuration and - as stated above in section 2.1. - the sky over Westeros seems to show only one. But contrary to the usual planets in binary stars, one needs only the light of one star to get the chaotic sequence of seasons in the Sitnikov problem! As long as there are two big objects acting on the planet with their gravity, it is not important if only one of them produces light. If the two stars started out with different masses, one of them will have ended its life before the other. It could then have exploded as a supernova and been reduced to a neutron star or black hole and thus not be visible on the sky of Westeros (see Figure 3). If the explosion has taken place during a phase, where the planet was far away from the orbital plane, the effects on the planet would have been minimal[6]. The light of the remaining star is sufficient to produce the seasonal effects that arise from the oscillatory motion of the planet as described above. Although it seems very unlikely and almost impossible for such a special configuration of stars and planets to form in a natural way and to exist in the real universe, we can (and will!) in that case simply refer to magic. If you don't believe in magic just consider the sheer number of potential planetary systems in a universe of roughly two trillion galaxies (Conselice et al, 2016) containing on average 200 billion stars each. There simply must be a Sitnikov system out there.

## 5. Conclusions

We have shown that from the many explanations brought forward by other authors to explain the seasons in the world of "Game of Thrones", none is able to satisfyingly address all details. There is only one astronomical explanation that works and that is the celestial mechanical case of the Sitnikov problem. Two stars and a planet that orbits along a straight line perpendicular to the orbital plane of the stars can build a planetary system, that has all the properties to reproduce any desired sequence of seasons. Not only that, but due to the nonlinear properties of the Sitnikov configuration and the mathematical theorem of Moser it is proven, that there is no sequence of seasons, however strange and obscure, that can not be found on a planet in such a system.

We thus have solved once and for all the problems surrounding the question if and when "Winter is coming".

*Acknowledgements*

We acknowledge the support of Christoph Lhotka, Karl Wodnar and Rudolf Dvorak and their extensive analysis of the Sitnikov Problem. Any errors and scientific inaccuracies are due to us (or more precisely: to the evil magical influence of the White Walkers, who do not want us to discover their secrets).

*References*

- Alekseev, V. (1968a). Quasirandom dynamical systems I. Math. USSR Sbornik, 5:73–128.
- Alekseev, V. (1968b). Quasirandom dynamical systems II. Math. USSR Sbornik, 6:505–560.
- Alekseev, V. (1969). Quasirandom dynamical systems III. Math. USSR Sbornik, 7:1–43.
- Conselice, C. et al. (2016). The Evolution of Galaxy Number Density at z < 8 and its Implications. Astrophysical Journal 830:17
- Dvorak, R. (1984). Numerical experiments on planetary orbits in double stars. Celestial Mechanics, 34:369–378.
- Dvorak, R. and Freistetter, F. (2005). Orbit Dynamics, Stability and Chaos in Planetary Systems. Lecture Notes in Physics 683: 3–140.
- Dvorak, R. and Lhotka, C. (2013). Celestial Dynamics. WILEY Berlin. ISBN: 978-3-527-40977-8
- Dvorak, R. and Lhotka. C. (2014). Scholarpedia, 9(12):11096., http://bit.ly/DvorakLhotka2014

---

[6] Maybe the influence was just right and localised enough to have caused the mystery phenomenon that is only known as the "doom of Valyria".

- Dvorsky, G. (2012). 5 Scientific Explanations for Game of Thrones' Messed-Up Seasons. io9. http://bit.ly/Dvorsky2012
- Griffith, P. and Douglas, T. (2017). Game of Thrones: There's a Scientific Reason for Westeros's Years-Long Winters. Vanity Fair. http://bit.ly/GriffithDouglas2017
- Kostov, V. et al (2013). Winter is coming. arXiv:1304.044. http://bit.ly/Kostov2013
- Laskar, J. et al (1993). Stabilization of the Earth's obliquity by the Moon. Nature 36: 615–617
- Lhotka, C (2015). Sitnikov's planet. UMI VIII: 1-31 . http://bit.ly/Lhotka2015
- MacMillan, W. (1911). An integrable case in the restricted problem of three bodies. Astron. J., 27:11–13.
- Merrifield, M. (2014). Why winter is coming. Deep Sky Videos. http://bit.ly/Merrifield2014
- Moser, J. (1973). Stable and Random Motion in Dynamical Systems, Annals of Mathematical Studies 77.
- Pilat Lohinger, E. et al. (2002). Stability of planetary orbits in double stars. Proceedings of the First European Workshop on Exo-Astrobiology. 547–548.
- Selcke, D. (2015). Game of Thrones Theorycraft: the reasons for the (weird) seasons. Winteriscoming.net. http://bit.ly/Selcke2015
- Sitnikov, K. (1961). The Existence of Oscillatory Motions in the Three-Body Problem. Soviet Physics Doklady, 5:647.
- Welsh, W. et al (2012). The Transiting Circumbinary Planets Kepler-34 and Kepler-35. Nature volume 481, pages 475–479.
- Wodnar, K. (1993). The original Sitnikov article. New insights. Celestial Mechanics and Dynamical Astronomy, 56:99–101.